\documentclass[prd,nofootinbib,twocolumn]{revtex4}
\usepackage{graphics}
\usepackage{bm}

\begin{document}

\title{
Islands in the $\Lambda$-sea: An alternative cosmological model
}

\author{
Sourish Dutta and Tanmay Vachaspati
}
\affiliation{
CERCA, Department of Physics, Case Western Reserve University,
10900 Euclid Avenue, Cleveland, OH 44106-7079, USA.}

\begin{abstract}
\noindent
We propose an alternate cosmological model in which our observable
universe is an island in a cosmological constant sea. Initially the 
universe is filled with cosmological constant of the currently observed 
value but is otherwise empty. In this eternal or semi-eternal de Sitter 
spacetime, we show that local quantum fluctuations (upheavals) can 
violate the null energy condition and create islands of matter. The 
perturbation spectra of quantum fields other than that responsible 
for the upheaval, are shown to be scale invariant. With further cosmic 
evolution the island disappears and the local universe returns to its 
initial cosmological constant dominated state.
\end{abstract}

\maketitle

\section{Introduction}
\label{introduction}

The last two decades have seen giant strides in observational
cosmology. We now have accurate characterization of the cosmic
large-scale structure, the cosmic microwave background radiation,
and the energy budget of the universe. In addition we have strong 
support for non-baryonic dark matter, and tantalizing evidence for 
dark energy. 

Theoretically the observations fit quite well into a cosmological 
framework in which a period of inflation in the earliest moments of 
cosmic history is postulated \cite{Guth:1980zm}. The driver for inflation 
is a scalar field called the ``inflaton'' and the dynamics of the
scalar field is governed by a potential function. Although 
there is considerable freedom to choose the scalar field content,
the potential function, and initial conditions, a large 
number of inflationary models have been constructed that agree 
very well with the whole slew of observations. Most spectacular of 
these is the general agreement of the WMAP data with a scale invariant 
distribution of adiabatic density fluctuations \cite{Peiris:2003ff}. 

\begin{figure}
\scalebox{0.40}{\includegraphics{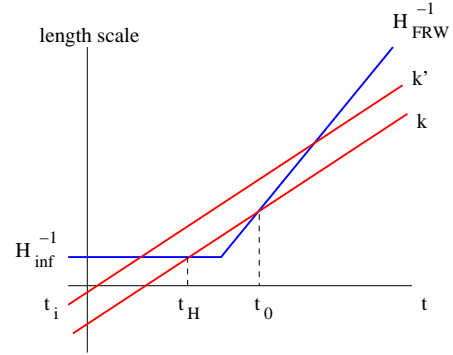}}
\caption{\label{fig1} Sketch of the behavior of the Hubble length 
scale with cosmic time in inflationary cosmology, and the evolution
of fluctuation modes. At early times, the vacuum energy is large and 
so the Hubble length scale is small and constant. Once the vacuum
energy decays, inflation ends, the universe reheats and enters FRW 
expansion. Then the Hubble length scale grows linearly with time. 
The physical wavelength of a fluctuation mode starts out less than 
$H^{-1}_{\rm inf}$ at some early time $t_i$. The mode is stretched to 
scales larger than the horizon at some time $t_H(k)$. The mode later
renters the horizon at some epoch $t_0 (k)$. For illustrative purposes 
we have depicted that the wavelength grows linearly with cosmic time. 
In reality, the wavelength grows in proportion to the scale factor.
}
\end{figure}

To understand the inflationary model and the generation of density
fluctuations, we plot the evolution of the Hubble length scale 
($H^{-1}$) in Fig. \ref{fig1}. The Hubble scale during inflation 
is a very small constant, typically of the order of $10^3 l_P$ where 
$l_P \sim 10^{-33}$ cm is the Planck length. After inflation ends, 
cosmological reheating occurs in which the universe gets filled with 
radiation. From then on the universe evolves as in a 
Friedmann-Robertson-Walker (FRW) cosmology. Density fluctuations arise
because there are quantum fluctuations in the inflationary period
whose wavelengths grow during the inflationary period and eventually 
become larger than the Hubble scale. Once the quantum modes are 
superhorizon, their amplitudes are frozen. During the FRW epoch,
these modes re-enter the horizon, and can give rise to the density
fluctuations required for large-scale structure formation.

\begin{figure}
\scalebox{0.40}{\includegraphics{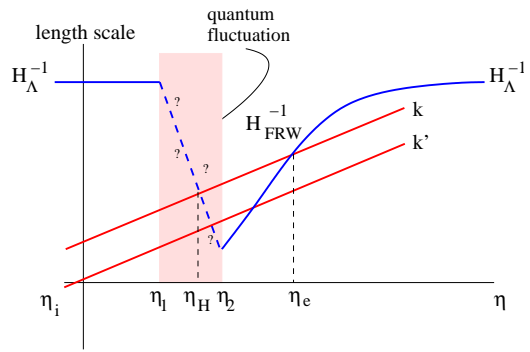}}
\caption{\label{fig2} Sketch of the behavior of the Hubble length 
scale with conformal time, $\eta$, in our cosmological model, and 
the evolution of fluctuation modes. At early times, inflation is 
driven by the presently observed dark energy, assumed to be a cosmological 
constant. As the cosmological constant is very small, the Hubble length
scale is very large -- of order the present horizon size. 
Exponential inflation in some horizon volume ends not due to the
decay of the vacuum energy as in inflationary scenarios but due
to a quantum fluctuation in the time interval $(\eta_1, \eta_2)$ 
that violates the null energy condition (NEC). The NEC violating
quantum fluctuation  causes the Hubble length scale to decrease. 
After the fluctuation is over, the universe enters radiation dominated 
FRW expansion, and the Hubble length scale grows with time. The 
physical wavelength of a quantum fluctuation mode starts out less 
than $H^{-1}_\Lambda$ at some early time $\eta_i$. The mode exits 
the cosmological horizon during the NEC violating fluctuation 
($\eta_H$) and then re-enters the horizon at some later epoch 
$\eta_e$ during the FRW epoch.
}
\end{figure}

In this paper, we propose a modified cosmology which is based on
a different hypothesis and investigate its viability. 
Instead of the evolution of the Hubble length scale shown in
Fig.~\ref{fig1} we consider the evolution depicted in Fig.~\ref{fig2}. 
We now give a short overview of the idea, explaining each stage of
the model. 
\begin{enumerate}
\item In the beginning the universe is inflating due to the observed 
dark energy \cite{Riess:1998cb,Perlmutter:1998np} that we assume is a 
cosmological constant ($\Lambda$)\footnote{We make no attempt to address 
why $\Lambda$ is so small compared to the Planck scale.}.  The de Sitter 
horizon size ($H^{-1}_\Lambda$) is comparable to our present horizon, 
($H^{-1}_0$).
\item A quantum fluctuation of some field ({\it e.g.} scalar field, 
photon) in a horizon-size volume in the expanding phase of de Sitter 
spacetime drives the Hubble constant to a large value. Even as the 
Hubble length scale is decreasing the universe continues to expand.
Such explosive events necessarily need to violate 
the null energy condition (NEC). We will show, following 
Refs.~\cite{GutVacWin??,Winitzki:2001fc,Vachaspati:2003de},
that quantum field theoretic fluctuations allow for this possibility.
\item After the NEC violating fluctuation is over, the Hubble constant 
is large and classical radiation fills the volume. Rapid interactions 
thermalize the radiation. This part of the universe then evolves as a 
radiation dominated FRW universe and we call it an ``island''.
\item With further evolution, the radiation in the FRW universe
dilutes and eventually the volume is again dominated by the
cosmological constant and the spacetime returns to its normal
inflating state. 
\end{enumerate}

Our idea has elements of earlier work on eternal inflation 
\cite{Vilenkin:1983xq,Linde:1986fd} and especially Garriga and Vilenkin's  
``recycling universe'' \cite{Garriga:1997ef} (see also the discussion
in \cite{Carroll:2004pn}). In eternal inflation scenarios,
quantum fluctuations in the inflaton field drive the Hubble length scale
to smaller values. In our model we also consider a quantum fluctuation
but it can occur in {\em any} quantum field and it has to be large. In 
both inflationary cosmology and our case, the quantum fluctuation needs 
to violate the NEC. Furthermore, in both cases the back-reaction of the 
fluctuation is assumed to lead to a faster rate of cosmological expansion. 
In the language of \cite{Dyson:2002pf}, the evolution we are considering
is one of the ``miraculous'' trajectories that go directly from a
dead de Sitter region of spacetime to a region that is ``macroscopically
indistinguishable from our universe'' (MIFOU). Eventually the 
trajectory leaves the MIFOU region and returns to the dead de Sitter
region. Our idea also has elements of Steady State Cosmology 
\cite{BonGol48,Hoy48} in which matter is sporadically produced by 
explosive events in a hypothetical C-field but spacetime is eternal. 
In our case, the explosive events are quantum field theoretic and 
produce an entire cosmos worth of matter. The ekpyrotic cosmological
model \cite{Khoury:2001wf}, like our model, also utilizes a decreasing 
Hubble length scale. In the ekpyrotic model, the decrease in the Hubble 
length scale is due to extra-dimensional brane-world physics and results 
in a a period of contraction of our three dimensional universe. In the 
present model though, the Hubble length scale decreases due to an
``ordinary'' quantum field theoretic fluctuation, but the universe 
continues to expand during the contraction of the Hubble length scale.

Having summarized the main idea, we now discuss each step of
this model in greater detail. In Sec.~\ref{pertspectrum}
we discuss a key test of the model  -- whether it predicts a scale 
invariant distribution of density fluctuations.

\section{NEC violations in de Sitter space}
\label{necindS}

The first stage of the model only relies on the presence of a 
cosmological constant. Observations indicate some form of dark 
energy and they are consistent with a cosmological constant. 
One feature of the first stage of our model is that it does not 
necessarily begin in a singularity -- there may be no big bang 
and spacetime need not be created out of nothing as in quantum
cosmology. All that we need is an expanding de Sitter background
and this can be part of a classical de Sitter spacetime with 
no beginning and no end, with early contraction and then expansion. 
We will only consider the expanding phase of the de Sitter spacetime 
in the following discussion. The scale factor of the universe at 
this stage is given by:
\begin{equation}
a(t) = a_0 e^{H_\Lambda t} \equiv - \frac{1}{H_\Lambda \eta}
\label{aoftdS}
\end{equation}
where $\eta \in (-\infty ,0)$ is the conformal time. 

In de Sitter spacetime, as well as any other spacetime, there are
fluctuations of the energy-momentum tensor, $T_{\mu \nu}$, of quantum 
fields. This follows simply due to the fact that the vacuum, $|0\rangle$,
is an eigenstate of the Hamiltonian but not of the energy-momentum
density operator, ${\hat T}_{\mu\nu}$. In short-hand notation: 
\begin{eqnarray}
{\hat T}_{\mu\nu} |0\rangle &=& \sum 
           [(\dots )a_l a^\dag_k + (\dots )a^\dag_l a^\dag_k] ~ |0\rangle
\nonumber \\
 &=& \sum [(\dots ) |0\rangle + (\dots ) |2;k,l \rangle]
\label{Tonvac}
\end{eqnarray}
where, the ellipses within parenthesis denote various combinations of
mode functions and their derivatives; $a^\dag_k$, $a_l$ are creation 
and annihilation operators and $|2;k,l\rangle$ is a two particle state. 
Since the final expression is
not proportional to $|0\rangle$, the vacuum is not an eigenstate of
${\hat T}_{\mu\nu}$ and there will be fluctuations of the energy-momentum
tensor in de Sitter space.

It has been shown 
\cite{GutVacWin??,Winitzki:2001fc,Vachaspati:2003de}
that quantum field theory of a light scalar field in the Bunch-Davies 
vacuum \cite{Bunch:1978yq} in de Sitter space leads to violations of 
the NEC. For the present application, we only need the general arguments 
of Refs.~\cite{GutVacWin??,Winitzki:2001fc,Vachaspati:2003de}
and not the detailed calculations. First, one constructs the 
``smeared NEC operator'' 
\begin{equation}
\hat{O}^{\rm ren}_{W}\equiv 
 \int d^{4}x\sqrt{-g}\, W\left( x; R,T\right) 
N^\mu N^\nu {\hat T}^{\rm ren}_{\mu\nu}
\label{OrenW}
\end{equation}
where $ W\left( x; R,T\right)$ is a smearing function 
on a length scale $R$ and time scale $T$. The vector
$N^\mu$ is chosen to be null, and the superscript 
$ren$ denotes that the operator has been suitably renormalized.
By the argument given below Eq.~(\ref{Tonvac}) we find that 
${\hat O}^{\rm ren}_W$ will fluctuate. On dimensional grounds:
\begin{equation}
O_{\rm rms}^2 \equiv 
\langle 0 | ({\hat O}^{\rm ren}_W)^2 | 0 \rangle \sim H_\Lambda^8
\label{Oren2expec}
\end{equation}
in the special case when $R = T = H_\Lambda^{-1}$. 
Since, in de Sittter space,
$\langle 0|{\hat T}_{\mu \nu} |0\rangle \propto g_{\mu\nu}$,
we also have:
\begin{equation}
\langle 0 | {\hat O}^{\rm ren}_W |0\rangle =0
\label{Orenexpec}
\end{equation}
Therefore the fluctuations of ${\hat O}^{\rm ren}_W$ are both positive and
negative. Assuming a symmetric distribution, we come to the conclusion
that quantum fluctuations of a scalar field violate the NEC with
50\% probability. Exactly the same arguments can be applied to 
quantum fluctuations of a massless gauge field such as the photon. 

The calculation described above shows that the NEC will be violated
by quantum fluctuations with 50\% frequency but does not give us the 
probability distribution of the violation amplitude. For that we would 
need to calculate the actual probability distribution for the operator 
${\hat O}^{\rm ren}_W$. However, by continuity we can expect that large 
amplitude NEC violations will also occur with some diminished but
non-zero probability.

\section{Extent and duration of NEC violation}
\label{extentandduration}

\begin{figure}
\scalebox{0.60}{\includegraphics{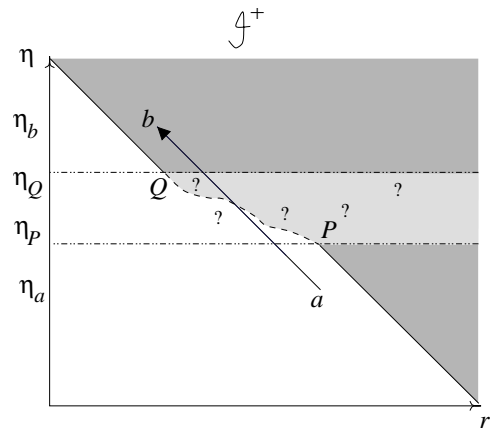}}
\caption{\label{nec_violation} 
We show a classical de Sitter spacetime for conformal time
$\eta < \eta_P$, that transitions to a faster expanding 
classical de Sitter spacetime for $\eta > \eta_Q$.
The inverse Hubble size is shown by the white region.
A bundle of ingoing null rays originating at point $a$ is 
convergent initially but becomes divergent in the superhorizon 
region at point $b$. This can only occur if the NEC
is violated in the region $\eta \in (\eta_P, \eta_Q)$. 
In the quantum domain, a classical picture of spacetime may 
not be valid and this is made explicit by the question marks. 
}
\end{figure}

There are NEC violating quantum fluctuations on all
spatial and temporal scales. However, most of these 
fluctuations are irrelevant for cosmology -- the spacetime
might respond very locally, and then return to its orginal 
state. As we now argue, only fluctuations that occur
on large spatial scales can have a lasting effect on the 
spacetime {\it i.e.} the faster expansion can continue
in a predictable way even after the NEC violating fluctuation 
is over. 

Consider the spacetime diagram of Fig.~\ref{nec_violation}.
In that diagram we show an initial de Sitter space that
later has a patch in which the space is again de Sitter 
though with a larger expansion rate. Hence the initial
Hubble length scale $H_i^{-1}$ is larger than the final
Hubble length scale $H_f^{-1}$. Therefore there are ingoing
null rays that are within the horizon initially that propagate
and are eventually outside the horizon. An example of such
a null ray is the line from $a$ to $b$. At point $a$ a bundle
of such rays will be converging whereas at point $b$ the
bundle will be diverging. The transition from convergence to
divergence of a bundle of null rays can only occur if there
is NEC violation somewhere along the null ray provided some
mild conditions are satisfied. This follows from the Raychaudhuri
equation. In the spherically symmetric case, the mild conditions 
are satisfied and hence the transition to faster expansion 
requires NEC violation. 

\begin{figure}
\scalebox{0.60}{\includegraphics{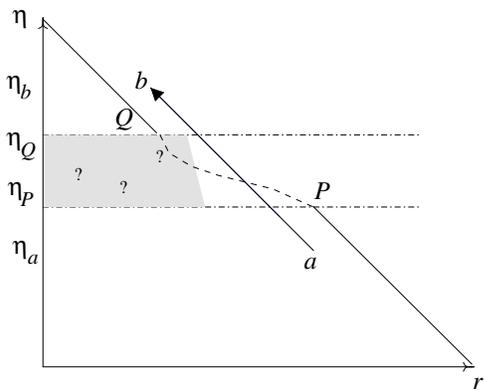}}
\caption{\label{nec_patch} 
A spacetime diagram similar to that in Fig.~\ref{nec_violation} 
but one in which the NEC violation occurs over a sub-horizon region
(shaded region in the diagram).
Now the null ray bundle from $a$ to $b$ goes from being converging 
(within the horizon) to diverging (outside the horizon). However, it 
does not encounter any NEC violation along its path, and this is not 
possible as can be seen from the Raychaudhuri equation. 
Since the ingoing null rays are convergent as far out as the point 
$P$, the size of the quantum domain has to extend out to at
least the inverse Hubble size of the initial de Sitter space.
Therefore the NEC violating patch has to extend beyond the initial 
horizon.
}
\end{figure}

Now we argue that the NEC violation has to extend over a region
that is at least as large as $H_i^{-1}$, if the faster expansion
is to last longer than the duration of the NEC violation\footnote{This
is similar to the argument in Ref.~\cite{Vachaspati:1998dy}
showing that inflation requires homogeneity on superhorizon 
scales as an initial condition.}. If NEC violation only occurred on 
a scale smaller than $H_i^{-1}$, one could imagine a null ray that would
never enter the NEC violating region and yet go from being converging
to diverging (see Fig.~\ref{nec_patch}). This would be inconsistent
with the Raychaudhuri equation.

If an NEC violation does occur in a sub-horizon region, it
could cause a temporary change in the expansion rate of the
region. When the NEC violating fluctuation in the small region 
of space ceases, the faster expanding local region would have to 
revert to the ambient expansion, or else some spacetime feature, 
such as a singularity, will have to occur that could prevent the 
traversal of a null ray from the slower expanding exterior region 
to the faster expanding interior region. Additional boundary conditions 
would need to be imposed at the singularity for the spacetime evolution 
to be predictable. An example of such a process can be found in 
Ref.~\cite{Borde:1998wa} in connection with topological inflation
\cite{Vilenkin:1994pv,Linde:1994hy}.

An additional intuitive argument can be given that might help understand 
the need for a large spatial region where NEC is violated. Whenever a 
faster expanding universe is created, it must be connected by a wormhole 
to the ambient slower expanding region. The wormhole can be kept open 
if the energy conditions are violated \cite{Morris:1988cz}. But, if the 
wormhole neck is small, as soon as the energy condition violations
are over, it must collapse and pinch off into a singularity. Signals 
from the singularity can propagate into the faster expanding universe and 
predictability will be lost. However, if the neck of the wormhole 
is larger than the horizon size of the ambient universe, the ambient 
expansion can hold up the wormhole and the neck does not collapse
even after the NEC violation is over.

Our argument that NEC violations on scales larger than
the horizon are needed to produce a faster expanding universe is 
consistent with earlier work \cite{Farhi:1986ty} showing that it is 
not possible to produce a universe in a laboratory without an initial 
singularity (also see \cite{Linde:1991sk}). Subsequent discussion of 
this problem in the quantum 
context \cite{Fischler:1989se,Farhi:1989yr,Fischler:1990pk}, however, 
showed that a universe may tunnel from nothing
without an initial singularity, just as in quantum cosmology
\cite{Vilenkin:1982de,Hartle:1983ai}. In this case,
the created universe is disconnected from the ambient $\Lambda$-sea.
Without an inflaton, the process would therefore produce a second
$\Lambda$-sea which would be empty until matter-producing fluctuations
of the kind we have been discussing can create islands.

Guided by these arguments, we conjecture that small regions with 
finite duration NEC violation cannot lead to predictable, faster 
expanding islands. The conjecture is clearly not proven but we 
believe there is substantial evidence in favor of it.
Hence we conclude that, to get a faster expanding region that 
lasts beyond the duration of the quantum fluctuation and remains
predictable, the spatial extent of the NEC violating fluctuation 
must be larger than 
$H_i^{-1}$: 
\begin{equation}
R > H_i^{-1}
\end{equation}
where $R$ is the spatial extent of the fluctuation and shows
up as the spatial smearing scale in the calculation of ${\hat O}_{\rm rms}$.

An explicit evaluation shows that ${\hat O}_{\rm rms}$ is proportional
to inverse powers of the temporal smearing scale and diverges as the 
smearing time scale $T \rightarrow 0$. Hence the briefer the fluctuation, 
the stronger it can be, as we might also expect from an application of 
the Heisenberg time-energy uncertainty relation. Therefore we
will take the time scale of the NEC violation to be vanishingly
small:
\begin{equation}
T \rightarrow 0
\end{equation}
In our analysis of the spectrum of density fluctuations in
Sec.~\ref{pertspectrum} we will refer to this as the ``sudden'' 
approximation.

\section{Backreaction on spacetime: a working hypothesis}
\label{backreaction}

The situation at hand has quantum fields but a classical
spacetime. Generally this situation is handled using
semiclassical relativity:
\begin{equation}
G_{\mu\nu} = 8\pi G \langle {\hat T}_{\mu\nu}^{\rm ren} \rangle
\end{equation}
where $\langle \rangle$ denotes the expectation value in some 
specified quantum state. Hence the spacetime in this formalism
only responds to the expectation value of the energy-momentum
tensor and fluctuations about the mean do not play any role.
However we are interested in precisely the effects
of fluctuations of ${\hat T}_{\mu\nu}^{\rm ren}$ and so it is 
essential that we go beyond semiclassical relativity
to be able to treat the backreaction of the NEC violating
fluctuations on the spacetime. For small fluctuations, one
could envisage expanding the metric around a fixed background
and quantizing the metric fluctuations. Such an attempt has been 
made in Refs.~\cite{Tsamis:1996qm} though not in the context
of NEC violations. The perturbative scheme can not however
hope to capture the physics of large fluctuations of the kind
we are interested in.

Since a rigorous treatment of the backreaction is not
possible, we shall adopt a ``working hypothesis'' in which
the NEC violating fluctuation behaves like ``phantom energy'' 
{\it i.e.} a classical perfect fluid with equation of state 
$w \equiv p/\rho < -1$ where $\rho > 0$. Furthermore, in the
sudden approximation discussed in the previous section, the
phantom energy exists only for a vanishingly small time period.
Hence the energy content of the universe has the following
time dependence:
\begin{eqnarray}
\rho = \Lambda \ , \ \ w = -1 \ , \ \ \eta < \eta_f \nonumber \\
\rho = \rho_{\rm FRW} \ , \ \ w=+\frac{1}{3} \ , \ \  \eta > \eta_f 
\label{rhow}
\end{eqnarray}
where $\eta_f$ denotes the instant at which the NEC violating
fluctuation occurs and $\rho_{\rm FRW}$ denotes the energy density 
after the NEC violating fluctuation is over and this is assumed to
be dominantly in the form of radiation. The initial condition for 
the FRW phase is: $\rho_{\rm FRW} (\eta_f) = \rho_{\rm mg}$,
the radiation density required for matter-genesis.

With this working hypothesis for the energy content of the local universe, 
the backreaction on the spacetime is given by:
\begin{equation}
H^2 = \frac{8\pi G}{3} \rho
\label{FRWeq}
\end{equation}
where, as usual, $H = {\dot a}/a$ and $a(t)$ is the local scale factor. 
For $\eta < \eta_f$, $\rho$ is a constant and $H= H_\Lambda$ which is
constant. In a vanishingly small interval around $\eta_f$, $\rho$ increases 
rapidly due to the NEC violating fluctuation and this means that
$H$ also increases correspondingly. This implies that the scale factor 
grows faster than exponentially during the NEC violating fluctuation,
yielding a vanishingly short period of ``super-inflation''. After
the NEC violating fluctuation is over, the region is filled with
radiation energy density and the FRW epoch starts. We summarize the
behavior of the Hubble scale, $H$, as follows:
\begin{eqnarray}
H = H_\Lambda \ , \  \ \eta < \eta_f \nonumber \\
H = H_{\rm FRW} \ ,  \ \  \eta > \eta_f 
\label{Hbehavior}
\end{eqnarray}
with the initial condition $H_{\rm FRW} (\eta_f) = H_{mg}$ where
$H_{mg}$ is the Hubble constant at the epoch of matter-genesis.

In writing Eq.~(\ref{FRWeq}), we are assuming that spacelike
surfaces of constant $\rho$ are flat. That is why we have not
included the spatial curvature term, $k/a^2$. This is consistent
with our working hypothesis since the initial state ($\eta < \eta_f$)
is flat and the universe expands even faster during the phantom
energy stage which is entirely classical. In the oft studied example 
where a scalar field tunnels through to a different value, it is 
known that the surfaces of constant field have negative spatial 
curvature. The energy density in the field is purely due to potential 
energy and so the surfaces of constant field are also surfaces of 
constant energy density. Hence the tunneling event produces an open 
universe with negative spatial curvature \cite{Bucher:1994gb}. The 
scenario in this paper is different from the tunneling scenario because 
there is no instanton that describes the NEC violating fluctuation.
It can be shown explicitly that the tunneling process preserves
de Sitter invariance \cite{Vachaspati:1991tq} (though see the 
caveat mentioned in Footnote~33 in \cite{Vilenkin:1994rn}) and 
this symmetry implies hyperbolic spatial slicings ({\it i.e.} open 
universe slicings) of the spacetime. In our case we know that NEC 
violations only occur if de Sitter invariance is broken. This can be 
seen by considering
\begin{equation}
\langle {\hat T}_{\mu\nu}^{\rm ren} {\hat T}_{\lambda\sigma}^{\rm ren} \rangle
\end{equation}
If we demand that this be a tensor respecting the de Sitter
symmetry, then it must be expressible in terms of the metric
tensor since this is the only tensor available to us. However,
then, when we contract with null vectors to get 
$\langle (N^\mu N^\nu {\hat T}_{\mu \nu}^{\rm ren} )^2 \rangle$, 
the result will be zero since $g_{\alpha \beta}N^\alpha N^\beta =0$,
and there will be no NEC violating fluctuations. 

The smearing process that is used in the calculation of $O_{\rm rms}$
is essential in quantum field theory since quantum operators are
distributions that are defined solely by their actions on test 
functions, much like the Dirac delta function (see Chapter 3 of
\cite{StrWig64}). Therefore only the smeared operator 
${\hat O}_W^{\rm ren}$ has physical meaning and it explicitly breaks 
de Sitter invariance. The fluctuations of ${\hat O}_W^{\rm ren}$ depend 
on the spacetime volume under consideration. 

In the preceding sections we have described the nature of the 
fluctuation and a working hypothesis for determining the backreaction 
of the fluctuation on the spacetime. We now discuss the likelihood of
getting the described NEC violating fluctuation.

\section{Likelihood -- the role of the observer}
\label{anthropics}

In Sec.~\ref{extentandduration} we have seen that a cosmologically 
relevant NEC violation must occur on superhorizon scales and will
most likely be on very short time scales. Even among such fluctuations,
most of the NEC violating fluctuations will be small in magnitude. 
The energy density produced as a result of small fluctuations will 
also be small and will not impact observational cosmology. However, 
there is a tiny probability of having a large fluctuation in 
which there is sufficient energy density to heat a horizon-size 
volume to a very high temperature. If the temperature produced is 
high enough, matter-genesis will occur in this region eventually 
leading to large-scale structure and to the cosmology we observe. If 
the temperature is not high enough for matter-genesis, the island
of energy will not lead to a matter dominated cosmology of the kind 
we know. We are assuming that matter-genesis is essential for
observers to exist. 

A large NEC violating fluctuation in which an island of matter is 
produced is very improbable. However, since spacetime is eternal 
in this model we can wait indefinitely for the island of matter to 
be produced. The island of matter can then proceed to thermalize, 
cool and form our present day cosmic environment. We also assume that 
the temperature required for magnetic monopoles production is higher 
than that required for matter-genesis. If the temperature at the 
beginning of the FRW phase is below that needed for monopole
formation but above the matter-genesis temperature then there will 
be no cosmological magnetic monopole problem. This solution to the 
monopole over-abundance problem is similar in spirit to that proposed 
in Ref.~\cite{Dvali:1995cj} where the Grand Unified phase transition
never occurs.

The probability of fluctuations in the $\Lambda$-sea that can lead to
an inflating cosmology versus those that produce an FRW universe have 
been considered by several researchers \cite{Dyson:2002pf,Albrecht:2004ke}.
In particular, Dyson et al. \cite{Dyson:2002pf} estimate probabilities based on 
a ``causal patch'' picture, with the conclusion that it is much more probable 
to directly create a universe like ours than to arrive at our present state via
inflation. Albrecht and Sorbo \cite{Albrecht:2004ke} have argued that the 
conclusion rests crucially on the causal patch picture, and provide a 
different calculation leading to the conclusion that inflationary cosmology 
is favored. Both calculations assume the existence of fields that are suitable 
for inflation. In this case, one should also include the additional possibility 
of ``creating a universe from nothing'' in the evaluation of relative probabilities.

Our hypothesis is that there exists no field that is suitable to be
an inflaton. So the comparison of the likelihood of inflation versus no 
inflation is moot. What is still relevant is the relative probability of 
obtaining an island which is different from ours but those in which observers
could live. We do not know the various necessary conditions for life.
However, once we have assumed that the end point of the NEC violating
fluctuation is a thermal state with all the different forms of matter
in thermal equilibrium, further evolution of the island simply follows
that of standard big bang cosmology.

In island cosmology, we need also ask where we are located on the
island. Are we close to the edge of the island (``beach'')? In that case 
we would observe anisotropies in the cosmic microwave background 
since in some directions we
would see the $\Lambda$-sea while in others we would see inland. 
However, the island is very large (by a factor $a_0/a_f$) compared to 
our present horizon, $H_\Lambda^{-1}$. If we assume a uniform probability 
for our location on the island, our distance from the $\Lambda$-sea will 
be an $O(1)$ fraction of $H_\Lambda^{-1} a_0/a_f$. Since $a_0 /a_f$ is 
of order $T_{mg}/T_0$ -- the ratio of the matter-genesis temperature
to the present temperature --  we are most likely to be sufficiently 
inland so as not to observe any anisotropy in the cosmic microwave
background.

Whereas inflationary models crucially rely on the existence of a 
suitable scalar field (inflaton), we have so far not specified 
the quantum field that causes the NEC violating fluctuation. 
We now turn to this issue.

\section{The NEC violating field}
\label{whichfield}

The phantom energy that is assumed to describe the effects of the
NEC violating quantum fluctuation, by definition, satisfies $\rho +p < 0$.
In addition, the assumption that the backreaction is given by 
Eq.~(\ref{FRWeq}), requires $\rho > 0$. Hence we need a quantum
field that can give NEC violating fluctuations while still having 
positive energy density. In other words, the energy density should
be positive but the pressure should be sufficiently negative so that
the NEC is violated.

First consider a scalar field, $\phi$, with potential $V(\phi )$. The
energy density and pressure are:
\begin{eqnarray}
{\hat \rho} &=& \frac{1}{2} {\dot \phi}^2 + \frac{1}{2} ({\bm \nabla} \phi )^2 
        + V(\phi ) \nonumber \\ 
{\hat p} &=& \frac{1}{2} {\dot \phi}^2 - \frac{1}{6} ({\bm \nabla} \phi )^2 
        - V(\phi ) 
\end{eqnarray}
where the hats on $\rho$ and $p$ emphasize that these are quantum
operators Therefore:
\begin{equation}
{\hat \rho} + {\hat p} = {\dot \phi}^2 + \frac{({\bm \nabla} \phi )^2}{3} 
\end{equation}

The operators ${\hat \rho}$ and ${\hat \rho}+{\hat p}$ are not
proportional to each other and fluctuations in one do not have
to be correlated with fluctuations of the other. The energy 
density in a region can be positive while the NEC is violated.
Therefore a scalar field, even if $V(\phi ) =0$, can provide suitable 
NEC violating fluctuations.

The particle physics in the very early stages of the model is
described by low energy particle physics that we know so well.
At present we do not have any experimental evidence for a scalar 
field. One field that we know of today is the electromagnetic 
field. Could the electromagnetic field give rise to a suitable NEC
violating fluctuation?

For the electromagnetic field we have:
\begin{eqnarray}
{\hat \rho} = \frac{1}{2} ( {\bf E}^2 + {\bf B}^2 ) \nonumber \\ 
{\hat p} = \frac{1}{6} ( {\bf E}^2 + {\bf B}^2 ) = \frac{1}{3} {\hat \rho}
\end{eqnarray}
So now ${\hat \rho}$ and ${\hat p}$ are not independent operators and
\begin{equation}
{\hat \rho} + {\hat p} = \frac{4}{3} {\hat \rho}
\end{equation}
From this relationship between the operators, it is clear that
the only electromagnetic fluctuation that can violate the NEC also 
has negative energy density. This means that even though the 
electromagnetic field can violate the NEC, it does not satisfy 
the positive energy density condition needed in the working hypothesis 
to find the backreaction. It may be possible that the electromagnetic 
field will still be found to be suitable once we know better how to 
handle the backreaction problem. Then perhaps we will not need to
rely on the working hypothesis that requires positive energy density.

There is a possible loophole in our discussion of the electromagnetic
field. The equation of state ${\hat p} = {\hat \rho}/3$ follows from the
conformal invariance of the electromagnetic field ${\hat T}^\mu_\mu =0$.
However, we know that quantum effects in curved spacetime give rise
to a conformal anomaly and the trace $\langle {\hat T}^\mu_\mu \rangle$ 
is not precisely zero. So we can expect that the equation of state 
${\hat p} = {\hat \rho}/3$ is also anomalous. Whether this anomaly 
can allow for NEC violations with positive energy density is not clear 
to us.

Note that it is not necessary for the NEC violation to originate from 
a fluctuation of a massless or light field. The arguments of 
Sec.~\ref{necindS} are very general and apply to massive fields 
as well.  Though, for a massive field, ${\hat O}_{\rm rms}$ will be 
further suppressed by exponential factors whose exponent depends 
on powers of $H_\Lambda /m$. While the likelihood of a suitable NEC 
violating flucutation from a very massive field is much smaller compared 
to that of a light or massless field, the massive field fluctuations 
are clearly more important if the light field doesn't even exist! The 
discussion in the previous section of the likelihood still applies.

\section{Spectrum of perturbations}
\label{pertspectrum}

In contrast to inflationary models, we do not have a classical field 
that is slowly rolling on some potential. Instead a mode of a field
(call it $\phi$) is undergoing an NEC violating quantum fluctuation. 
In general there will also be quantum fluctuations of the other modes 
of the field and these will give rise to density fluctuations. 
In addition to fluctuations of $\phi$, there will be other fields 
that will undergo quantum fluctuations in the rapidly changing 
spacetime and these will also give rise to density fluctuations. 
Some of these will be massless or light compared to $H_\Lambda$
and for others the mass will be important. However, we might
expect that the mass will be important only if it is larger than
the {\it final} value of the Hubble parameter after the fluctuation 
is over, denoted by 
\begin{equation}
H_f \equiv H (\eta_f+)
\end{equation} 
We will not pursue massive fields further but restrict our attention 
to massless fields for the present.

First we will consider a light field other than $\phi$, and not
interacting directly with $\phi$,
and find its power spectrum. Let us denote such a field 
generically by $\chi$, its eigenmodes by 
$\chi_k (\eta ) \exp(i{\bf k}\cdot {\bf x})$ and, as is
commonly done in the theory of cosmological perturbations
\cite{Mukhanov:1990me}, define the variable 
\begin{equation}
v_k (\eta) \equiv a(\eta ) \chi_k (\eta )
\end{equation}
with $k = | {\bf k} |$.
If $\chi$ is a massless, minimally coupled scalar field then 
$v_k$ satisfies:
\begin{equation}
v_k'' + \biggl ( k^2 - \frac{a''}{a} \biggr ) v_k = 0
\label{vkeq}
\end{equation}
where primes denote differentiation with respect to the
conformal time $\eta$ ($dt = a d\eta$). The solution of
Eq.~(\ref{vkeq}) with suitable initial conditions (described 
below) directly leads to the power spectrum of $\chi$
perturbations at any time via:
\begin{equation}
{\cal P}_\chi (k, \eta) = \frac{k^3}{2\pi^2} 
              \left | \frac{v_k}{a} \right |^2
\label{Pchik}
\end{equation}

In the de Sitter phase of the cosmology, {\it i.e.} for
$\eta < \eta_f$,
\begin{equation}
\frac{a''}{a} = \frac{2}{\eta^2}
\end{equation}
The exact solution of Eq.~(\ref{vkeq}) with the boundary condition
that small wavelength modes go over into Minkowski space modes is:
\begin{equation}
v_k = \frac{ e^{-ik\eta}}{\sqrt{2k}} 
               \left ( 1 - \frac{i}{k\eta} \right ) \ , \ \ \eta < \eta_f
\end{equation}
The other independent mode is $v_k^*$. These are the mode functions for 
the Bunch-Davies vacuum \cite{Bunch:1978yq}. A derivation of these
mode functions using inverse scattering technology can be found in the 
Appendix.

In the radiation dominated FRW epoch, {\it i.e.} for $\eta > \eta_f$,
we have 
\begin{equation}
a (t) = a_f \sqrt{\frac{t}{t_f}}
\end{equation}
where $t_f$ is the cosmic time at which the NEC violating fluctuation 
occurs, and $a_f \equiv a(t_f)$. (Note that the Hubble parameter is
discontinuous at $\eta_f$ in the sudden approximation but the scale factor
is continuous.) In terms of the conformal time one finds:
\begin{equation}
a(\eta ) = a_f + a_f^2 H_{f} (\eta - \eta_f)
\label{aeta}
\end{equation}
Clearly $a'' =0$. Therefore:
\begin{equation}
v_k = \alpha_k e^{-ik\tau} + \beta_k e^{+ik\tau} \ , \ \ 
      \tau \equiv \eta - \eta_f > 0
\end{equation}

Next we need to solve Eq.~(\ref{vkeq}) at $\eta =\eta_f$. This
step is non-trivial since $a$ is continuous at $\eta_f$
but $a'$ is discontinuous. Hence $a''$ has a delta function
contribution. Using Eqs.~(\ref{aoftdS}) and (\ref{aeta}) we find:
\begin{equation}
\frac{a''}{a} = \frac{2}{\eta^2} \Theta (\eta_f - \eta )
                + a_f \Delta H \delta (\eta -\eta_f)
\end{equation}
where $\Theta (\cdot )$ is the Heaviside function and 
\begin{equation}
\Delta H \equiv H_f - H_\Lambda \approx H_f
\end{equation} 

Integrating Eq.~(\ref{vkeq}) in an infinitesimal interval around
$\eta_f$, we find the junction conditions:
\begin{eqnarray}
v_k (\eta_f +) &=& v_k (\eta_f -) \nonumber \\
v_k' (\eta_f +)&=& v_k' (\eta_f -) + a_f \Delta H v_k (\eta_f)
\end{eqnarray}
where the last term is due to the $\delta -$function piece in
$a''/a$. 
We can now find the coefficients $\alpha_k$ and $\beta_k$
by inserting the de Sitter and FRW mode functions and their
derivatives at $\eta =\eta_f$ in the junction conditions.
This gives:
\begin{eqnarray}
\alpha_k = \frac{1}{2} \left [ v_{kf-} + 
          \frac{i}{k} \biggl ( v_{kf-}' + a_f H_f v_{kf-} \biggr )
                       \right ] \nonumber \\
\beta_k = \frac{1}{2} \left [ v_{kf-} - 
          \frac{i}{k} \biggl ( v_{kf-}' + a_f H_f v_{kf-} \biggr )
                       \right ] 
\label{akbk}
\end{eqnarray}
where $v_{kf-} \equiv v_k (\eta_f-)$ and similarly for the 
(conformal) time derivative $v_k'$.

%

We are interested in the long wavelength fluctuations for which
$k\eta_f \rightarrow 0$. Then the dominant contributions 
come from the $v_{kf-}'$ and $a_f H_f v_{kf-}$ terms in 
Eq.~(\ref{akbk}) and are of order $1/(k\eta_f )^2$. However, the
$a_f H_f v_{kf-}$ term is much larger than the $v_{kf-}$ term
because $H_f >> H_\Lambda$. (Recall from Eq.~(\ref{aoftdS}) that 
$a_f \eta_f = - 1/H_\Lambda$.) Therefore
\begin{eqnarray}
\alpha_k &\approx& + \frac{1}{2\sqrt{2k}} \frac{1}{(k\eta_f)^2} 
                                           \frac{H_f}{H_\Lambda}\nonumber \\
\beta_k &\approx& - \frac{1}{2\sqrt{2k}} \frac{1}{(k\eta_f)^2} 
                                           \frac{H_f}{H_\Lambda}
\end{eqnarray}
Therefore
\begin{equation}
v_k (\eta) \approx \frac{-i}{\sqrt{2k}} \frac{1}{(k\eta_f )^2} 
                       \frac{H_f}{H_\Lambda} \sin (k\eta)
\label{vketak}
\end{equation}
Using Eqs.~(\ref{aeta}) and (\ref{vketak}) in (\ref{Pchik}), 
together with $\eta_k >> \eta_f$ gives:
\begin{equation}
{\cal P}_\chi (k, \eta_k) \approx 
          \frac{1}{4\pi^2} \frac{1}{a_f^4 H_{\Lambda}^2 \eta_f^4}
          \biggl ( \frac{\sin (k\eta)}{k\eta} \biggr )^2
\end{equation}
Making use of Eq.~(\ref{aoftdS}), $a_f \eta_f = -1/ H_\Lambda$, 
and taking the limit $k\eta \rightarrow 0$, we finally get:
\begin{eqnarray}
{\cal P}_\chi (k, \eta_k) \approx \frac{H_\Lambda^2}{4\pi^2} 
\label{Pchikfinal}
\end{eqnarray}
Since the result does not depend on $k$, the spectrum of $\chi$
fluctuations is scale invariant, as in the inflationary case
\cite{MukChi81}, with amplitude set by the cosmological constant.

As discussed earlier in this section, the result in 
Eq.~(\ref{Pchikfinal}) applies to all very light or massless fields
other than, and not interacting directly with, the NEC violating field.
In particular, in the context of the gravitational wave power 
spectrum, the perturbation of the metric is equivalent 
to $\chi /m_P$ where $m_P$ is the Planck mass.  Hence the power in 
gravitational waves is proportional to $(H_\Lambda / m_P )^2$ and 
is very tiny.

We now turn to the problem of estimating fluctuations of the NEC 
violating field itself. The NEC violating field (called $\phi$) is 
decomposed in a part that is responsible for the fluctuation 
($\phi_0$) and another ($\delta \phi$) that takes into account 
additional small fluctuations. Hence we write:
\begin{equation}
\phi = \phi_0 + \delta \phi
\end{equation}
where, in contrast to the inflationary case, {\it both} $\phi_0$ and 
$\delta \phi$ are quantum operators. 

To calculate density fluctuations due to $\delta \phi$, one needs a 
suitable model for the evolution of $\phi_0$ during the NEC violating 
fluctuation. This evolution is quantum and not described as a solution
to some classical equation of motion. The closest related problems that 
have been addressed in the literature are the production of particles 
during the quantum creation of the universe and the fluctuations of 
a vacuum bubble that has itself been produced in a tunneling event 
\cite{Rubakov:1984pa,Vachaspati:1991tq,Garriga:1991tb}. These analyses 
rely on the existence of an instanton describing the tunneling event. 
In our case, the NEC violation is not described by an instanton; 
instead it is described by the most probable fluctuation leading
to matter-genesis. Hence the existing techniques do not apply 
directly and new techniques are needed. We leave this as an open 
problem for future work.

\section{Assumptions}
\label{parallels}

Island cosmology involves several assumptions that we 
have pointed out above but now summarize and discuss. 

Our first assumption is that the dark energy is a cosmological 
constant. This is consistent with observations and moreover is
the simplest explanation of the Hubble acceleration. We assume
that the cosmological constant provides us with a background 
de Sitter spacetime that is eternal\footnote{For a discussion 
of the timescale on which the spacetime can remain de Sitter, 
see Ref.~\cite{Goheer:2002vf}.}. As de Sitter spacetime also
has a contracting phase, the singularity theorems of
Ref.~\cite{Borde:2001nh} are evaded.

The second assumption is that there is a scalar field in the model 
responsible for the NEC violation. It would have been more 
satisfactory if the electromagnetic field could have played
this role but we have shown (up to the loophole of the conformal 
anomaly) that the conformal invariance of the electromagnetic field 
prevents NEC violations with positive energy density. It is possible
that with a better understanding of the backreaction of quantum
energy-momentum fluctuations on the spacetime, the electromagnetic 
field might still provide suitable NEC violations (see 
Sec.~\ref{backreaction}).
 
The basic formalism of quantum field theory in curved spacetime 
clearly leads to NEC violations and this is not an assumption.
(Though one could reasonably question the applicability of quantum
field theory on systems with horizons.) Then there seems little doubt 
that there should exist large amplitude NEC violations, though 
occurring much more infrequently than the small amplitude violations. 
The idea that NEC violating fluctuations could have played an important 
cosmological role is also to be found in the ``eternal inflation'' 
scenario \cite{Borde:1997pp}. Indeed, the current scenario may also be 
viewed as an eternal inflation scenario -- since the universe is eternally 
inflating due to a cosmological constant! While we may not be able to 
test the idea of cosmological NEC violating fluctuations, we can certainly 
test quantum fluctuations with and without horizons in laboratory 
experiments 
\cite{Unruh:1980cg,Fischer:2004bf,Jacobson:1998he,Vachaspati:2004wn}. 

The third assumption we have made has to do with NEC violations
in regions of small spatial extent. Based on work done on the possibility
of creating a universe in a laboratory, topological inflation, and
wormholes, we have argued for the conjecture that small scale violations 
of NEC can only give rise to universes that are affected by signals 
originating at a singularity. Hence predictability is lost in such 
universes. Our assumption is that even if we did know how to handle the 
spacetime singularities affecting these universes, they would turn
out to be unsuitable for matter genesis. Without this assumption, 
we should also be considering such universes as possible homes.

The fourth assumption is that the final state of the fluctuation is 
a thermal state. All the different energy components are
also assumed to be in thermal equilibrium. We have then assumed that the
critical temperature needed for observers to exist is the temperature
at which matter-genesis occurs. One could relax this assumption
but one would need an adequate characterization of the most likely state 
to be able to calculate cosmological observables (e.g. spectrum of density 
fluctuations).

Our fifth assumption is our ``working hypothesis'' for the 
backreaction of the NEC violating fluctuation. This seems to be the 
weakest assumption in our analysis. However, we cannot do any better 
at the moment because the backreaction of quantum fluctuations on the 
spacetime requires that we consider a quantum theory of gravity as 
well. The backreaction problem also occurs in eternal inflationary 
cosmology where a similar working hypothesis is used. It would
be worthwhile to address the issue of backreaction using
quantum cosmology, string theory, or loop gravity, but that
project is beyond the scope of this paper. 

This brings us to the part of the model where we argue that even 
if the large amplitude fluctuations are infrequent, they are the 
only ones that are relevant for observational cosmology. This is 
quite similar to the arguments given in the context of eternal
inflationary cosmology where thermalized regions are relatively 
rare but these are the only habitable ones. It also occurs in chaotic 
inflation \cite{Linde:1983gd}, where closed universes of all 
sizes and shapes are produced
but only a few are large and homogeneous enough to develop into the 
present universe. So this part of our model is no weaker (and harder 
to quantify) than other cosmological models.

\section{Conclusions}
\label{conclusions}

To conclude, we have investigated a new cosmological model, which
we call ``island cosmology'', where large NEC violating quantum 
fluctuations (``upheavals'') in a cosmological constant de Sitter 
universe create islands of matter. In island cosmology, spacetime
may be non-singular and eternal\footnote{The essential point is to 
have an expanding de Sitter phase; whether this is part of an eternal
de Sitter spacetime or originates at a big bang makes no difference.}.
We have shown that fields other than the NEC violating field yield 
a scale invariant spectrum of perturbations. 

The spectrum of density perturbations due to the NEC violating field 
itself is left as an open problem. Determining this spectrum will
be crucial to determining if the model agrees with observations.
For example, if the scale of fluctuations in this field is still 
set by $(H_\Lambda /m_P)^2$ then the fluctuations are too small to
seed the structure that we know and the island will be a desert.
On the other hand, if the scale is set by $(H_f/m_P)^2$ then
there is a chance that island cosmology can be a viable model.
In that case, quantum NEC violations provide a definite mechanism 
by which regions that are ``macroscopically indistinguishable
from our universe'' can be produced from the dead de Sitter sea.

Suppose the spectrum of density fluctuations in island cosmology
turns out to agree with observations. Then the question would
arise if we can somehow distinguish island cosmology from an
inflationary scenario that is also consistent with observations.
Unfortunately the answer seems to be that no cosmological 
observation can distinguish between the scenarios because
of the immense adaptability of inflationary models. The
only distinguishing feature would have to come from the field
theory side since island cosmology does not rely on constructing
a suitable potential for a scalar field whereas this seems
to be a crucial feature in inflationary models. If the electromagnetic 
field is subsequently determined to be capable of providing suitable NEC 
violations, scalar fields might be dispensed off entirely in
island cosmology. 

The flip side of island cosmology is that if the density fluctuations 
turn out not to agree with observations, we can dismiss the
scenario of our universe being an island in the $\Lambda$-sea,
even though quantum field theory predicts the existence of such
islands. This in itself would be an interesting conclusion.

\begin{acknowledgments} 
We would like to thank Jaume Garriga, Chris Gordon, Glenn Starkman,
Mark Trodden and, especially, Alex Vilenkin for discussion, and the 
Michigan Center for Theoretical Physics for hospitality. 
We are also grateful to Raphael Bousso, Sean Carroll, 
and Andrei Linde for comments. This work was supported by the 
U.S. Department of Energy and NASA.
\end{acknowledgments}

\appendix

\section{Solution for the de Sitter mode functions}
\label{dSmodesoln}

Here we will solve the mode function equation based on 
a technique encountered in the inverse scattering
literature (for example, see Ref.~\cite{Kwong:1985ti}).

The differential equation we wish to solve, Eq.~(\ref{vkeq}),
can be written as:
\begin{equation}
- v_k'' + \frac{a''}{a} v_k = k^2 v_k
\end{equation}
which is a Schrodinger equation with potential $a''/a$ and
eigenvalue $k^2$. Now
\begin{equation}
\frac{a''}{a} = f' + f^2
\end{equation}
where $f \equiv a'/a$. Therefore the Hamiltonian is:
\begin{equation}
H = -\partial^2 + f' + f^2 = (\partial + f) (-\partial +f)
\end{equation}
Denoting
\begin{equation}
Q = -\partial + f \ , \ \ Q^+ = \partial + f
\end{equation}
we can write:
\begin{equation}
H = Q^+ Q 
\end{equation}
Therefore the original differential equation is:
\begin{equation}
Q^+ Q v_k = k^2 v_k
\label{Qvkeq}
\end{equation}

Now consider the ``partner'' Hamiltonian:
\begin{equation}
H_- = Q Q^+ = - \partial^2 - f' + f^2
\end{equation}
and consider the partner Schrodinger equation:
\begin{equation}
Q Q^+ u_E = E u_E
\label{uEeq}
\end{equation}
{\it Claim:} if $u_E$ is a solution to Eq.~(\ref{uEeq}) then 
$v_k = Q^+ u_E$ is a solution to Eq.~(\ref{Qvkeq}) with $k^2 =E$. 
This claim is easy to check since: 
\begin{equation}
Hv_k = Q^+Q (Q^+u_E) = Q^+ H_- u_E = Ev_k
\end{equation}
This result is useful because if we know the eigenstates of the partner
Schrodinger equation, we can find the eigenstates of the original
Hamiltonian by applying $Q^+ = \partial + f$ to the partner eigenstate.

In the case of mode functions in de Sitter space ($a=- 1/H\eta$),
\begin{equation}
f = \frac{a'}{a} = - \frac{1}{\eta}
\end{equation}
and hence
\begin{equation}
H_- = - \partial^2 - f' +f^2 = -\partial^2
\end{equation}
Since this is the Hamiltonian for modes in Minkowski space, we 
observe that the de Sitter and Minkowski Hamiltonians are partners. 

The eigenstates of $H_-$ are:
\begin{equation}
u_k = \frac{C_k}{\sqrt{2k}} e^{-ik\eta} 
\end{equation}
where $C_k$ is a normalization factor, with $k$ being any real 
number. Therefore 
\begin{equation}
v_k = Q^+ u_k = (-ikC_k) \frac{e^{-ik\eta}}{\sqrt{2k}} 
                      \left ( 1 - \frac{i}{k\eta} \right )
\end{equation}
The correctly normalized mode function is obtained with $-ikC_k =1$.

This technique can be applied to the case of a massive 
scalar field too where the solutions are known in terms
of Hankel functions. While the solution is not as simple,
the technique enables us to find a relation between mode 
functions for two different parameters.

\end{document}